\documentclass[
  reprint,
  amsmath,amssymb,
  aps,
  prb,
  superscriptaddress
]{revtex4-2}

\usepackage{graphicx}
\usepackage{bm}
\usepackage{dcolumn}
\usepackage{comment}

\begin{document}

\title{Magnetotransport Properties of Iron-Based Ladder Compounds BaFe$_2$S$_3$ and BaFe$_2$Se$_3$ under High Pressure}

\author{Takuya Aoyama}
\email{aoyamatakuya@hiroshima-u.ac.jp}
\affiliation{
Program of Quantum Matter, Graduate School of Advanced Science and Engineering,
Hiroshima University, Hiroshima 739-8530, Japan
}
\affiliation{
International Institute for Sustainability with Knotted Chiral Meta Matter (WPI-SKCM2),
Hiroshima University, Hiroshima 739-8531, Japan
}

\author{Kenya Ohgushi}
\affiliation{
Department of Physics, Graduate School of Science,
Tohoku University, Sendai 980-8578, Japan
}

\begin{abstract}
The magnetotransport properties of the iron-based ladder compounds
BaFe$_2$S$_3$ and BaFe$_2$Se$_3$ under various pressures were investigated.
BaFe$_2$S$_3$ had a $T^2$ dependence in electrical resistivity and a normal
magnetoresistance near an insulator-metal transition.
Therefore, it had Fermi-liquid-like properties in the normal phase.
However, BaFe$_2$Se$_3$ had a $T^{3/2}$ dependence in electrical resistivity
and a negative magnetoresistance.
This suggested that BaFe$_2$Se$_3$ had non-Fermi-liquid properties near the
insulator-metal transition.
For both materials, the Hall resistivities indicated a hole-dominant magnetic
field dependence.
During a pressure-induced insulator-metal transition, BaFe$_2$Se$_3$ had a
rapid decrease in electrical resistivity above 15 GPa attributed to an
isomorphic structural transition previously reported in X-ray diffraction
experiments.
This transition had a strong impact on the electronic properties.
\end{abstract}

\maketitle

\section{Introduction}
    Iron-based ladder compounds $A$Fe$_2$$X_3$ ($A$ = Ba, Cs, $X$ = S and Se) exhibit Mott insulating states at ambient pressure because their low-dimensional crystal structures have a large $U$/$W$, where $U$ is on-site Coulomb repulsion and $W$ is the single-electron bandwidth~\cite{Caron2011,Caron2012,Nambu2012, Du2012,Mourigal2015,Hawai2015,Hawai2017}.
    BaFe$_2$$X_3$ ($X$ = S and Se) compounds undergo an insulator-metal Mott transition under pressure and become superconducting ~\cite{Takahashi2015,Yamauchi2015,Ying2017}.
    Therefore, BaFe$_2$$X_3$ ($X$ = S and Se) are favorable experimental platforms to continuously access an antiferromagnetic insulating phase, a weakly correlated metallic phase, and a superconducting phase in multi-orbital systems.

    The origin of these electronic states comes from the unique crystal structure of BaFe$_2$$X_3$ ($X$ = S and Se).
    Specifically, the Fe$^{2+}$ ions are tetrahedrally coordinated by $X$ ions, and local electronic structures are expected to be similar to those of iron-based superconductors, such as BaFe$_2$As$_2$ and FeSe.
    However, a global crystal network is formed by introducing a periodic deficiency of Fe atoms in the two-dimensional square lattice.
    Electron motion is restricted within the ladder, and therefore the ground state becomes an antiferromagnetic insulator, in stark contrast to the antiferromagnetic semimetal of iron-based superconductors.

\begin{figure}
        \centering
        \includegraphics[width=9cm,pagebox=cropbox,clip]{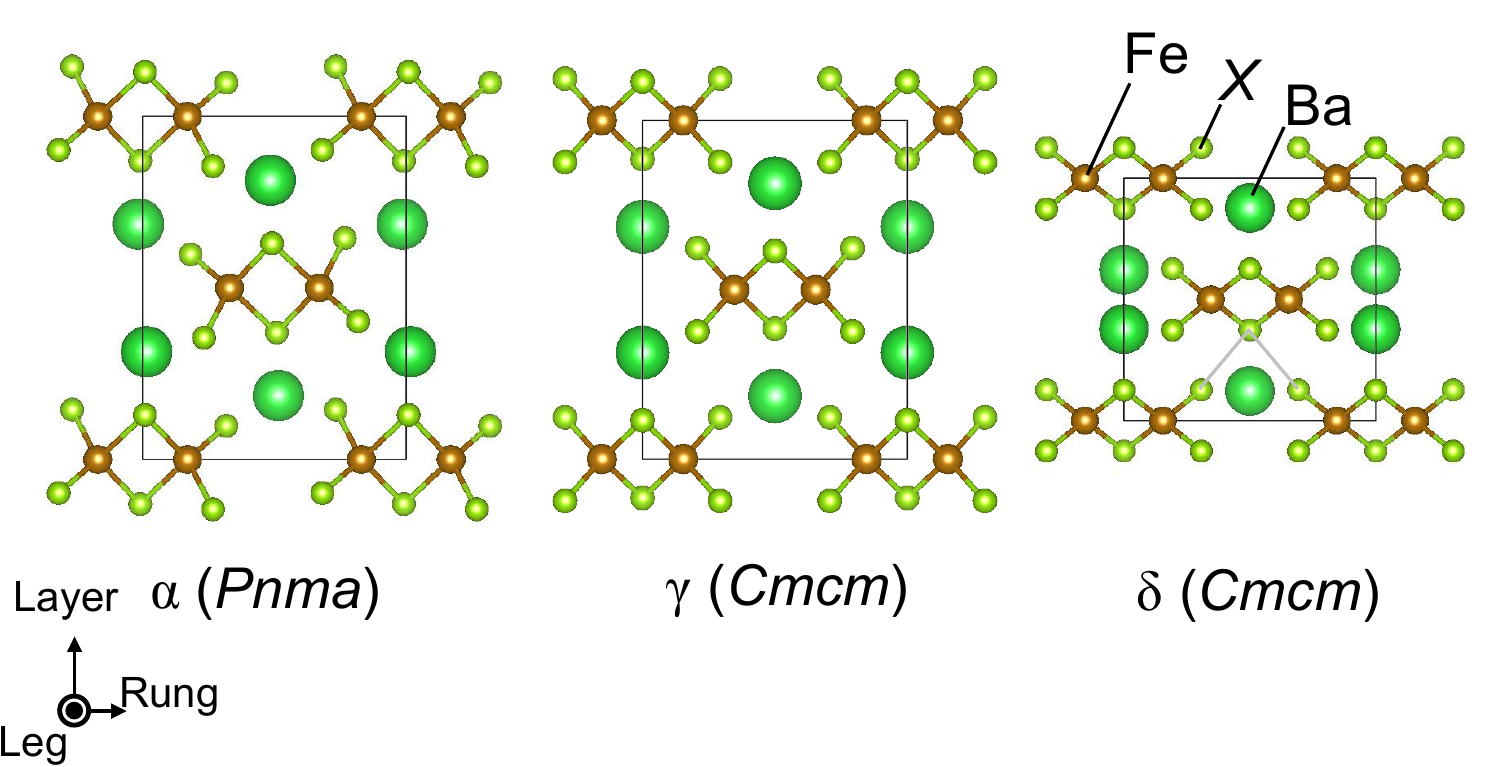}
        \caption{(color online) Crystal structures of BaFe$_2X_3$ ($X$ = S and Se) viewed from the leg direction. In the $\alpha$ structure, the ladder is buckled layer by layer, resulting in the $Pnma$ space group. The $\gamma$ structure has the highest symmetry ($Cmcm$). In the $\delta$ structure, the layer direction is greatly compressed, without any symmetry breaking from the space group of $Cmcm$.}
        \label{structure}
\end{figure}

    BaFe$_2$$X_3$ ($X$ = S and Se) compounds have a unique structural variation as shown in Fig. \ref{structure}.
    While the global structural motif of the two-leg ladder is maintained, several structures ($\alpha$, $\alpha^{\prime}$, $\gamma$, and $\delta$) with different symmetries are generated by buckling distortions between ladders and local distortions within the ladders, depending on composition, temperature and pressure.
    Here, we describe the details of $\alpha$, $\alpha^{\prime}$, $\gamma$, and $\delta$ structures~\cite{Hong1972, Svitlyk2013a, Svitlyk2019,Aoyama2019,Zheng2020}.
    The $\alpha$ structure adopts the CsAg$_2$I$_3$-type structure (space group $Pnma$), while $\alpha^{\prime}$ exhibits space group $Pmn2_1$ resulting from inversion symmetry breaking caused by Fe-Fe dimerization within the ladder.
    Both $\gamma$ and $\delta$ structures have CsCu$_2$Cl$_3$-type structures with the space group $Cmcm$; a key difference is that the $\delta$ structure is highly collapsed in the layer direction.
    The crystal structures strongly affect the orbital states of the iron. 
    The highly symmetric $\gamma$ structure has a uniform orbital arrangement, while the low-symmetry $\alpha$ and $\alpha^{\prime}$ structures have a staggered orbital arrangement~\cite{Imaizumi2020, Aoyama2023}.

    BaFe$_2$S$_3$ has the $\gamma$ structure at ambient pressure.
    In contrast, BaFe$_2$Se$_3$ exhibits a unique structural transition as a function of temperature and pressure.
    It has a $\alpha^{\prime}$ structure at room temperature under ambient pressure.
    With increasing temperature, it undergoes structural phase transitions $\alpha^{\prime}$ $\rightarrow$ $\alpha$ $\rightarrow$ $\gamma$ at 400 K and 660 K.
    As noted above, the crystal structures strongly affect the orbital states of the iron. 
    The highly symmetric $\gamma$ structure has a uniform orbital arrangement, while the low-symmetry $\alpha$ and $\alpha^{\prime}$ structures have a staggered orbital arrangement.
    At room temperature, sequential structural transitions from $\alpha^{\prime}$ $\rightarrow$ $\gamma$ $\rightarrow$ $\delta$ occur at pressures of 4 GPa and 15 GPa~\cite{Svitlyk2019}.
    The structural phase transition $\alpha^{\prime}$ $\rightarrow$ $\gamma$ at 4 GPa has been attributed to switching of orbital states~\cite{Aoyama2023}, while the origin of the isomorphic structural phase transition $\gamma$ $\rightarrow$ $\delta$ at 15 GPa is still experimentally unclear.    

    The magnetic structures of BaFe$_2$$X_3$ ($X$ = S and Se) have a variation that is closely related the crystal structure.
    Specifically, BaFe$_2$S$_3$ has a stripe magnetic order at 120 K under ambient pressure~\cite{Takahashi2015}.
    Under pressure, the stripe magnetic order is gradually suppressed and completely disappears near the insulator-metal transition, as revealed by $\mu$SR and M\"{o}ssbauer spectroscopy~\cite{Zheng2018, Materne2019}.
    In contrast, BaFe$_2$Se$_3$ has a block magnetic order under ambient pressure~\cite{Caron2012, Nambu2012}.
    Under 4 GPa pressure, that changes to the stripe magnetic order when the crystal structure changes from $\alpha^{\prime}$ to $\gamma$, as revealed by powder neutron diffraction~\cite{Wu2019,Zheng2022}. 
    At approximately 10 GPa, the insulator-metal transition occurs in BaFe$_2$Se$_3$, which then results in superconductivity at the phase boundary~\cite{Ying2017,Takahashi2022}.
    The presence of the stripe magnetic order adjacent to the superconducting phase is a common feature of BaFe$_2$S$_3$.
    In BaFe$_2$Se$_3$, however, the presence or absence of magnetic order in the metallic phase remains unclear.

    Here, we investigated the temperature dependence of electrical resistivity and magnetotransport properties of BaFe$_2$$X_3$ ($X$ = S and Se) under pressure to clarify the electronic states near the insulator-metal transition and the $\gamma \rightarrow \delta$ phase transition.
    In the magnetotransport measurements, there were significant differences between BaFe$_2$S$_3$ and BaFe$_2$Se$_3$.
    BaFe$_2$S$_3$ had a positive magnetoresistance in the normal state near the superconducting transition, suggesting normal magnetoresistance in the non-magnetic state.
    In contrast, BaFe$_2$Se$_3$ had a negative magnetoresistance in the normal state near the superconducting transition, suggesting residual magnetic degrees of freedom.
    In addition, its electrical resistivity decreased by an order of magnitude with the $\gamma \rightarrow \delta$ structure transition, and changed to a semimetallic phase with a small temperature dependence.

\section{Experimental}
    Single crystals of BaFe$_2$S$_3$ and BaFe$_2$Se$_3$ were grown via the slow cooling method.
    The stoichiometric amounts of Ba, Fe, S, and Se starting materials were placed in a carbon crucible in a glove box under an argon atmosphere.
    The crucible was placed inside a quartz tube that was then sealed under vacuum~\cite{Hirata2015, Imaizumi2020}.
    The quartz tube was placed in an electric furnace that was heated to 1423 K, and then cooled slowly to 923 K for 40 h.
    Under pressure, the electrical resistivity was measured via the four-terminal method, using a diamond anvil cell with a culet diameter of 400 $\mu$m.
    Inconel 625 insulated with alumina powder was used as a gasket and polyimide was used as a pressure-transmitting medium.
    The current was applied along the leg direction, and the pressure was determined by room-temperature ruby fluorescence.
    The low-temperature, high-field, high-pressure environment was realized by inserting the diamond anvil cell into a combined apparatus developed by Quantum Design.
    
\section{Results}
\subsection{BaFe$_2$S$_3$}
\subsubsection{Temperature dependence of electrical resistivity}
\begin{figure}[h]
    \centering
    \includegraphics[width=8cm,pagebox=cropbox,clip]{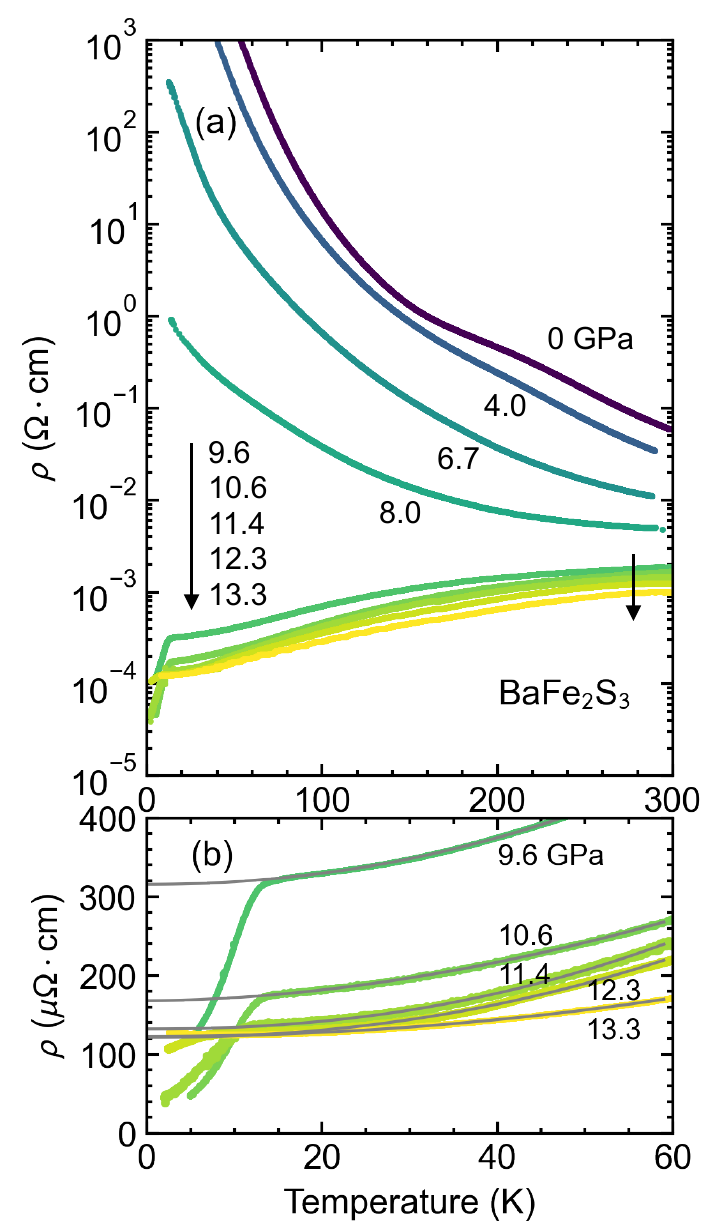}
    \caption{(color online) (a) Temperature dependence of BaFe$_2$S$_3$ electrical resistivity ($\rho$) under high pressures (log-scale). (b) Metallic temperature dependence of BaFe$_2$S$_3$ electrical resistivity above 9.6 GPa (linear scale). Solid lines are fits with the power law, $\rho$ = $\rho_0$ + $AT^n$.}
    \label{rhoT_BaFe2S3}
\end{figure}

    Figure \ref{rhoT_BaFe2S3}(a) shows the temperature dependence of electrical resistivity for BaFe$_2$S$_3$ under pressures ranging over $P$ = 0–13.3 GPa.
    At $P$ = 0 GPa, a broad hump associated with orbital ordering can be seen at $T$ $\sim$ 200 K~\cite{Hosoi2020, Takubo2025, Iwasaki2025}.
    The hump was suppressed by applying pressure and the $\rho$ had a monotonic increase with cooling at $P$ $\sim$ 6.7 GPa.
    The insulator-metal transition occurred at $P$ = 8–9.6 GPa, where the electrical resistivity at room temperature reached $\rho$ $\sim$ 10$^{-3}$ $\Omega\cdot$cm.
    This value approached the critical conductance of 1/$\rho$ $\sim$ 0.03$e^2/\hbar d$, where $e$, $d$, and $\hbar$ are electric charge, lattice constant, and Dirac$^{\prime}$s constant, respectively.
    These behaviors were consistent with previous results reported by Yamauchi $et$ $al$.~\cite{Yamauchi2015}.
\begin{figure}[h]
    \centering
    \includegraphics[width=6cm,pagebox=cropbox,clip]{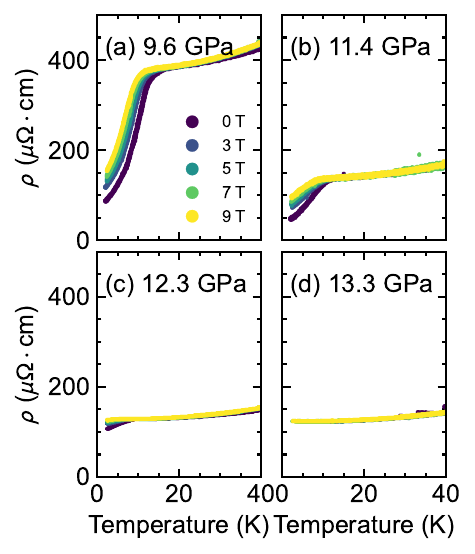}
    \caption{(color online) Temperature dependence of BaFe$_2$S$_3$ electrical resistivity ($\rho$) under several constant magnetic fields. 
        (a)-(d) Electrical resistivity $\rho$ measured at (a) 9.6 GPa, (b) 11.4 GPa, (c) 12.3 GPa, and (d) 13.3 GPa.
        Each panel displays data obtained under static magnetic fields up to 9 T applied perpendicular to the layer.}
    \label{TC_BaFe2S3}
\end{figure}

    At 9.6 GPa, $\rho$ had a metallic temperature dependence at all measured temperatures.
    An important feature was the steep drop in $\rho$ at $T_c$ at 10 K. 
    Because it was suppressed by the application of a magnetic field (Fig. \ref{TC_BaFe2S3}), this drop was likely attributed to the superconducting transition.
    The superconducting transition was observed at 9.6 GPa $\leq$ $P$ $\leq$ 12.3 GPa, and the maximum superconducting transition temperature was $T_c$ = 12.5 K at $P$ = 9.6 GPa.
    The superconducting phase disappeared at 13.3 GPa and $\rho$ had a monotonic temperature dependence down to the lowest temperature, indicating the loss of superconductivity.
    These behaviors were totally consistent with previous results reported by Yamauchi $et$ $al$, where a cubic anvil-type apparatus was used instead of a diamond anvil cell ~\cite{Yamauchi2015}.

\begin{figure}
    \centering
    \includegraphics[width=6cm,pagebox=cropbox,clip]{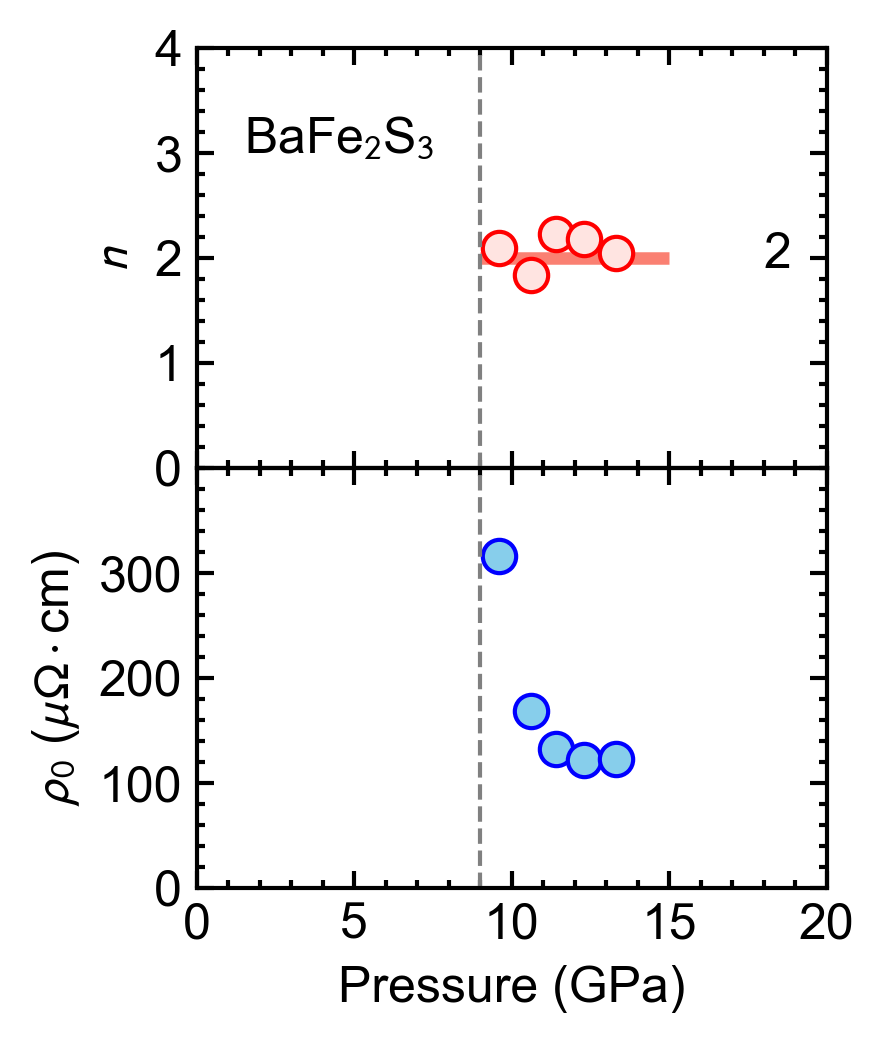}
    \caption{(color online) Pressure dependence of $n$ and $\rho_0$, where $n$ and $\rho_0$ are the exponent of $T$ for electrical resistivity and the residual resistivity in BaFe$_2$S$_3$. The vertical dotted line indicates the pressure of the insulator-metal transition.}
    \label{A_BaFe2S3}
\end{figure}

    To gain insights into the electronic state of the normal state around the insulator-metal transition, we investigated the power law of electrical resistivity, $\rho$ = $\rho_0$ + $A T^n$, where $\rho_0$ is a residual resistivity and $A$ is a coefficient of the temperature with power exponent $n$.
    The exponent $n$ in metals depends on the scattering mechanism. For example, acoustic phonons have $n$ = 5 far below the Debye temperature, electron-electron interactions have $n$ = 2, and antiferromagnetic spin fluctuations have $n$ = 3/2 in a three-dimensional system.
    The experimental data at 15 K $\leq$ $T$ $\leq$ 60 K could be fitted with the power law equation with $n$ = 2 in 9.6 $\leq$ $P$ $\leq$ 13.3 GPa (Fig. \ref{rhoT_BaFe2S3}(b)), indicating that the normal state near the superconducting phase was a Fermi liquid.
    Changes in $n$ and $\rho_0$ vs. pressure are shown in Fig.\ref{A_BaFe2S3}, which indicates that $\rho_0$ decreases with increasing $P$.

\subsubsection{Magnetotransport properties}
\begin{figure}[h]
    \centering
    \includegraphics[width=8cm,pagebox=cropbox,clip]{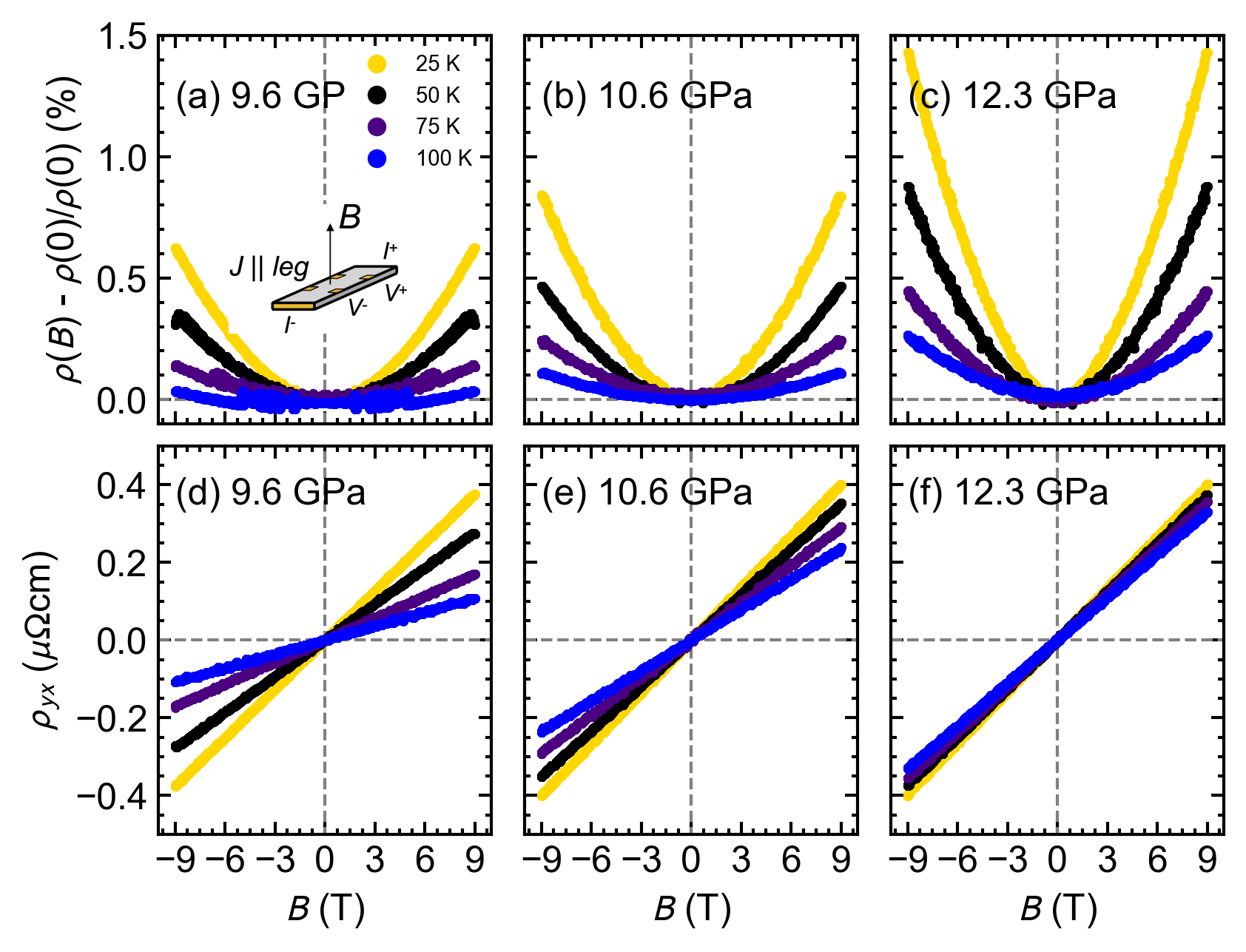}
    \caption{(color online) The magnetic field dependencies of (a-c) magnetoresistance and (d-f) Hall resistivity collected at (a, d) 9.6 GPa, (b, e) 10.6 GPa, and (c, f) 12.3 GPa for BaFe$_2$S$_3$. The inset in (a) illustrates the experimental configuration.}
    \label{rhoB_BaFe2S3}
\end{figure}

    Figure \ref{rhoB_BaFe2S3} plots the magnetoresistance and the Hall resistivity in the metallic phase of BaFe$_2$S$_3$ at various pressures. 
    The data were acquired in a transverse configuration with the current applied in the leg direction and the magnetic field applied along the layer direction (see inset of Fig.\ref{rhoB_BaFe2S3}(a)). 
    Above 9.6 GPa, which shows the metallic temperature dependence in resistivity, the resistivity increased in proportion to the square of the applied magnetic field, and its amplitude tended to increase at lower temperatures.
    This was a characteristic behavior of the normal magnetoresistance effect, which was enhanced by the pressure.
    The Hall resistivity shown in Fig. \ref{rhoB_BaFe2S3}(d-f) indicated that its dependence on the magnitude of the magnetic field was temperature dependent at 9.6 GPa, whereas the temperature dependence was suppressed under higher pressures. 
    Thus, applied pressure made the system more like a normal metal, which was consistent with the observation of a temperature-dependent electrical resistivity.

\begin{figure}[h]
        \centering
        \includegraphics[width=6cm,pagebox=cropbox,clip]{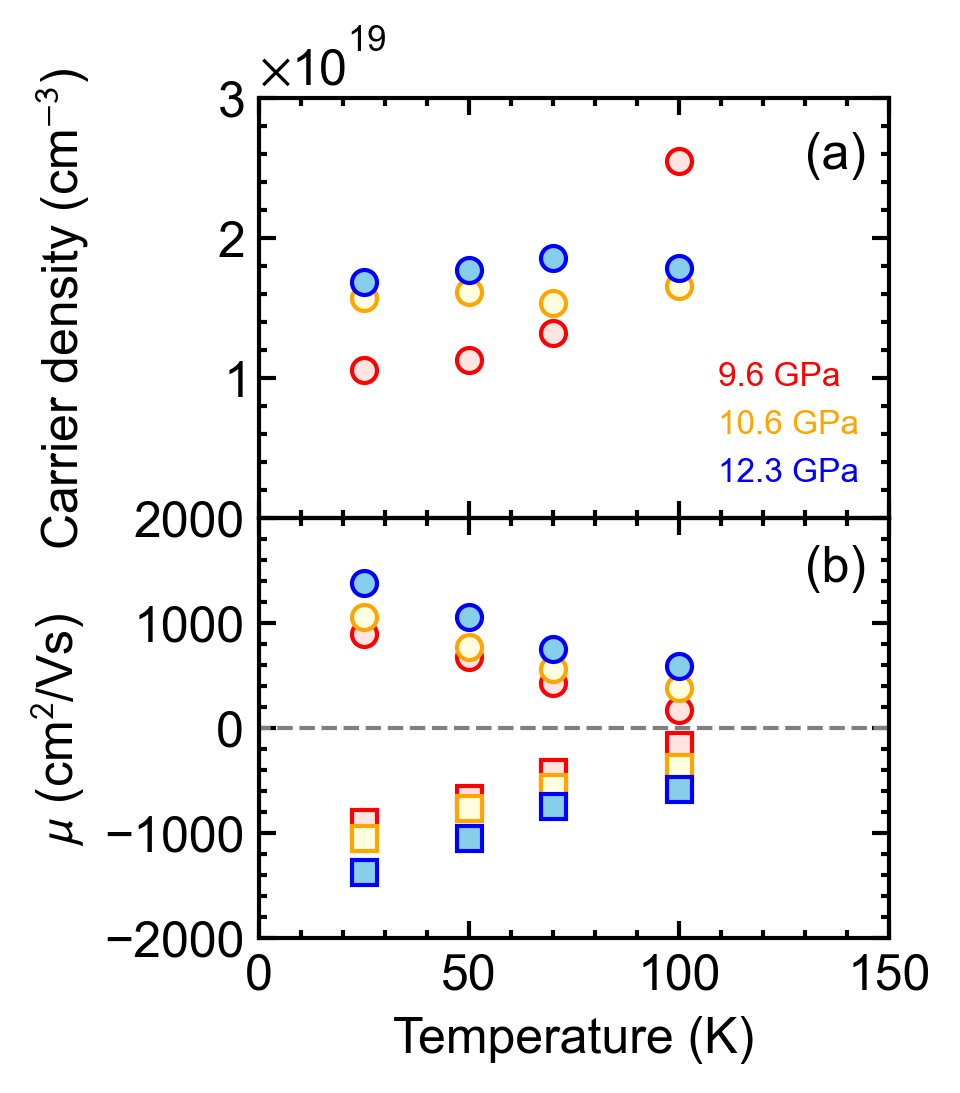}
        \caption{(color online) Temperature dependence of (a) carrier density and (b) signed mobility in BaFe$_2$S$_3$, estimated by using the two-carrier model. Positive and negative mobility values correspond to hole and electron carriers.}
        \label{mu_BaFe2S3}
    \end{figure}

    We interpret magnetotransport properties in the framework of the two-carrier model:
\begin{equation*}
    \begin{aligned}
    &\frac{\rho_{xx}(B)-\rho_{xx}(0)}{\rho_{xx}(B)} = \frac{q_1q_2n_1n_2\mu_1\mu_2(\mu_1-\mu_2)^2}{(q_1n_1\mu_1 + q_2n_2\mu_2)^2}B^2 \\[8pt]
    &\rho_{yx}(B) = \frac{q_1n_1\mu_1^2+q_2n_2\mu_2^2}{(q_1 n_1 \mu_1 + q_2 n_2 \mu_2)^2}B \\
    &\quad - \frac{q_1q_2n_1n_2\mu_1^2\mu_2^2(q_1n_1+q_2n_2)(\mu_1-\mu_2)^2}{(q_1 n_1 \mu_1 + q_2 n_2 \mu_2)^4}B^3
    \end{aligned}
\end{equation*}
    ,where $q_i$, $n_i$, and $\mu_i$ are the charge, carrier density, and signed mobility of the $i$-th carrier, respectively.
    Here, assuming that the two carriers have opposite signs ($q_1$ = $-q_2$ = $e$) and equal carrier density ($n_1$ = $n_2$ = $n$), we can obtain $\frac{\rho_{xx}(B)-\rho_{xx}(0)}{\rho_{xx}(B)}$  = -$\mu_1\mu_2 B^2$ and $\rho_{yx}(B) = \frac{\mu_1+\mu_2}{en(\mu_1-\mu_2)}B$.
    Temperature dependences of the carrier density and the mobility at several pressures are shown in Figs.~\ref{mu_BaFe2S3}(a) and (b). 
    In the vicinity of the insulator-metal transition, the carrier density was temperature-dependent. 
    With increasing pressure, the carrier density became independent of temperature.
    For both carriers, mobility increased with cooling and pressure.
    
    According to band calculations for the metallic phase, the 3$d_{zx}$ (3$d_{x^2-y^2}$) orbital forms electron (hole) pockets at the Fermi energy~\cite{Arita2015}.
    In addition, the renormalization factor $Z$, calculated using the slave-spin technique for the Hubbard model constructed from band calculations, indicated that the $d_{zx}$ orbital had a smaller $Z$ in the metallic phase, indicating that the electron pocket had a larger effective mass of quasi-particles~\cite{Pizarro2019}.
    These results suggest that hole carriers in 3$d_{x^2-y^2}$ orbitals with small effective masses contributed to the Hall effect, which explains the observed positive Hall coefficient.   

\subsection{BaFe$_2$Se$_3$}
\subsubsection{Temperature dependence}

\begin{figure}[h]
    \centering
    \includegraphics[width=8cm,pagebox=cropbox,clip]{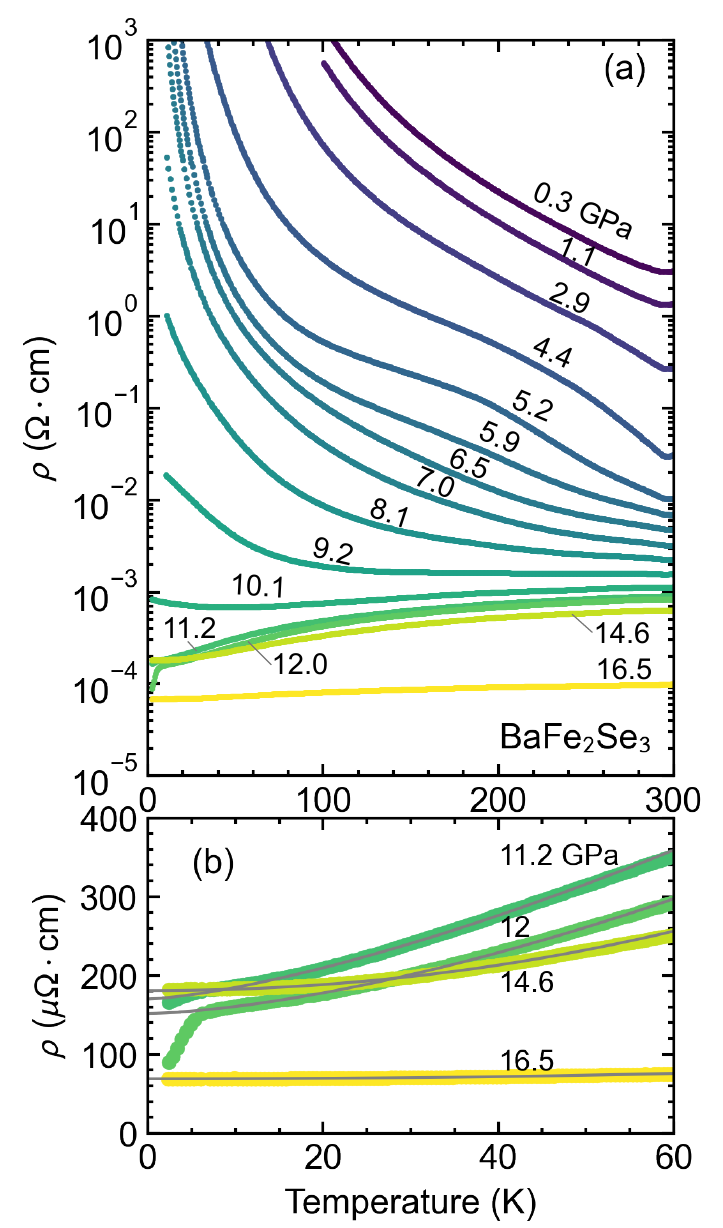}
    \caption{(color online) (a) Temperature dependence of BaFe$_2$Se$_3$ electrical resistivity ($\rho$) under high pressures (log-scale). (b) The metallic temperature dependence of its electrical resistivity above 11.2 GPa (linear scale). Solid lines show fits with the power law, $\rho$ = $\rho_0$ + $AT^n$.}
    \label{rhoT_BaFe2Se3}
\end{figure}

    Figure \ref{rhoT_BaFe2Se3}(a) shows the temperature dependence of the BaFe$_2$Se$_3$ electrical resistivity at pressures up to 16.5 GPa.
    At $P$ = 0 GPa, it was a Mott insulator because of predominant electron correlation effects, reflecting the low dimensionality of the ladder structure.
    Upon application of pressure $P$ = 4 GPa, the $T$-dependence of $\rho$ changed qualitatively from a simple activation-type behavior at 0 GPa $\leq$ $P$ $\leq$ 4 GPa to a slow divergence with a hump at 4.4 GPa $\leq$ $P$ $\leq$ 7 GPa.
    The latter was similar to the electrical resistivity of the $\gamma$ structure in BaFe$_2$S$_3$.
    Therefore, this could be interpreted that the structural transition from $\alpha^{\prime}$ to $\gamma$ structure occurred as reported previously ~\cite{Aoyama2023}.
    At higher pressures, the hump disappeared above 7.0 GPa and the electrical resistivity increased monotonically with lower temperatures.
    At 9.2 GPa, an insulator-metal transition occurred.

\begin{figure}[h]
    \centering
    \includegraphics[width=6cm,pagebox=cropbox,clip]{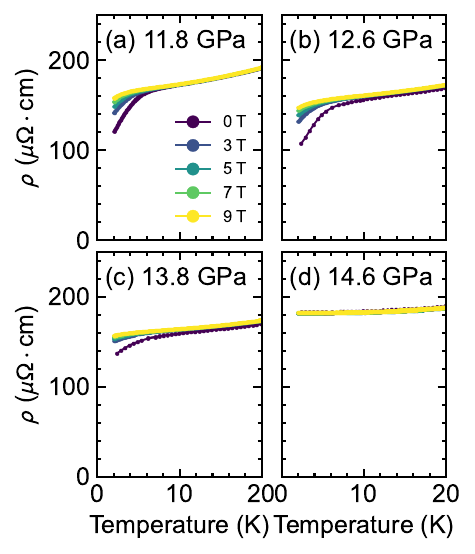}
    \caption{(color online) Temperature dependence of resistivity $\rho$ for BaFe$_2$Se$_3$ under various constant magnetic fields. 
        (a)-(d) Electrical resistivity $\rho$ measured at (a) 11.8 GPa, (b) 12.6 GPa, (c) 13.8 GPa, and (d) 14.6 GPa. 
        Each panel displays data obtained under static magnetic fields up to 9 T applied perpendicular to the layer.}
    \label{TC_BaFe2Se3}
\end{figure}

    Above $P$ = 10.1 GPa, $\rho$ had a metallic temperature dependence at all the measured temperatures.
    The steep drop of $\rho$ was observed at 12 GPa at 9 K.
    Because it was suppressed by the application of a magnetic field (Fig.~\ref{TC_BaFe2Se3}), we concluded that the drop was due to the superconducting transition.
    The superconducting transition was observed at 11.8 GPa $\leq$ $P$ $\leq$ 13.8 GPa, and the maximum superconducting transition temperature was $T_c$ = 5.0 K at $P$ = 11.8 GPa.
    At $P$ = 14.6 GPa, $\rho$ had a monotonic temperature dependence down to the lowest temperature, indicating the disappearance of superconductivity.
    These behaviors were basically consistent with previous results of Ying $et$ $al$, where a diamond anvil cell was used~\cite{Ying2017}.
    After the superconductivity disappeared upon pressurization, the electrical resistivity decreased remarkably over the entire temperature range for $P$ $>$ 16 GPa.
    This pressure agreed with that for the $\gamma$ $\rightarrow$ $\delta$ phase transition observed in X-ray diffraction measurements~\cite{Svitlyk2013a}, indicating that the $\gamma$ $\rightarrow$ $\delta$ phase transition was accompanied by a large change in the electronic state.
    
\begin{figure}[h]
    \centering
    \includegraphics[width=6cm,pagebox=cropbox,clip]{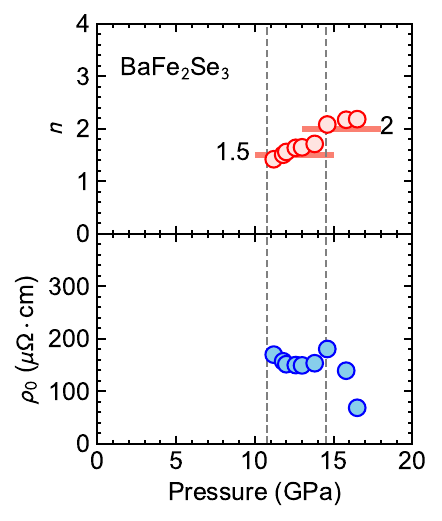}
    \caption{(color online) Pressure dependence of $n$ and $\rho_0$, where $n$ and $\rho_0$ are the exponent of $T$ for electrical resistivity and the residual resistivity in BaFe$_2$Se$_3$.The vertical dotted lines indicate the pressure of the insulator-metal transition and the lattice collapse transition.}
    \label{n_BaFe2Se3}
\end{figure}

    To evaluate the temperature dependence of electrical resistivity in the metallic region, the values above 11 GPa were fitted with $\rho$ = $\rho_0$ + $AT^n$ (Fig. \ref{rhoT_BaFe2Se3}(b)), and the temperature dependencies of $\rho_0$ and $n$ are shown in Fig. \ref{n_BaFe2Se3}.
    The exponent $n$ was close to 3/2 at 11.2 GPa near the superconducting transition, suggesting that three-dimensional antiferromagnetic fluctuations were dominant scattering sources.
    With increasing pressure, $n$ increased and then changed discontinuously to $n$ = 2 at 14.6 GPa, where the superconductivity disappeared, suggesting that the Fermi-liquid-like electronic state occurred in the $\delta$ phase.
    
\subsubsection{Magnetotransport properties}
\begin{figure}[h]
    \centering
    \includegraphics[width=8cm,pagebox=cropbox,clip]{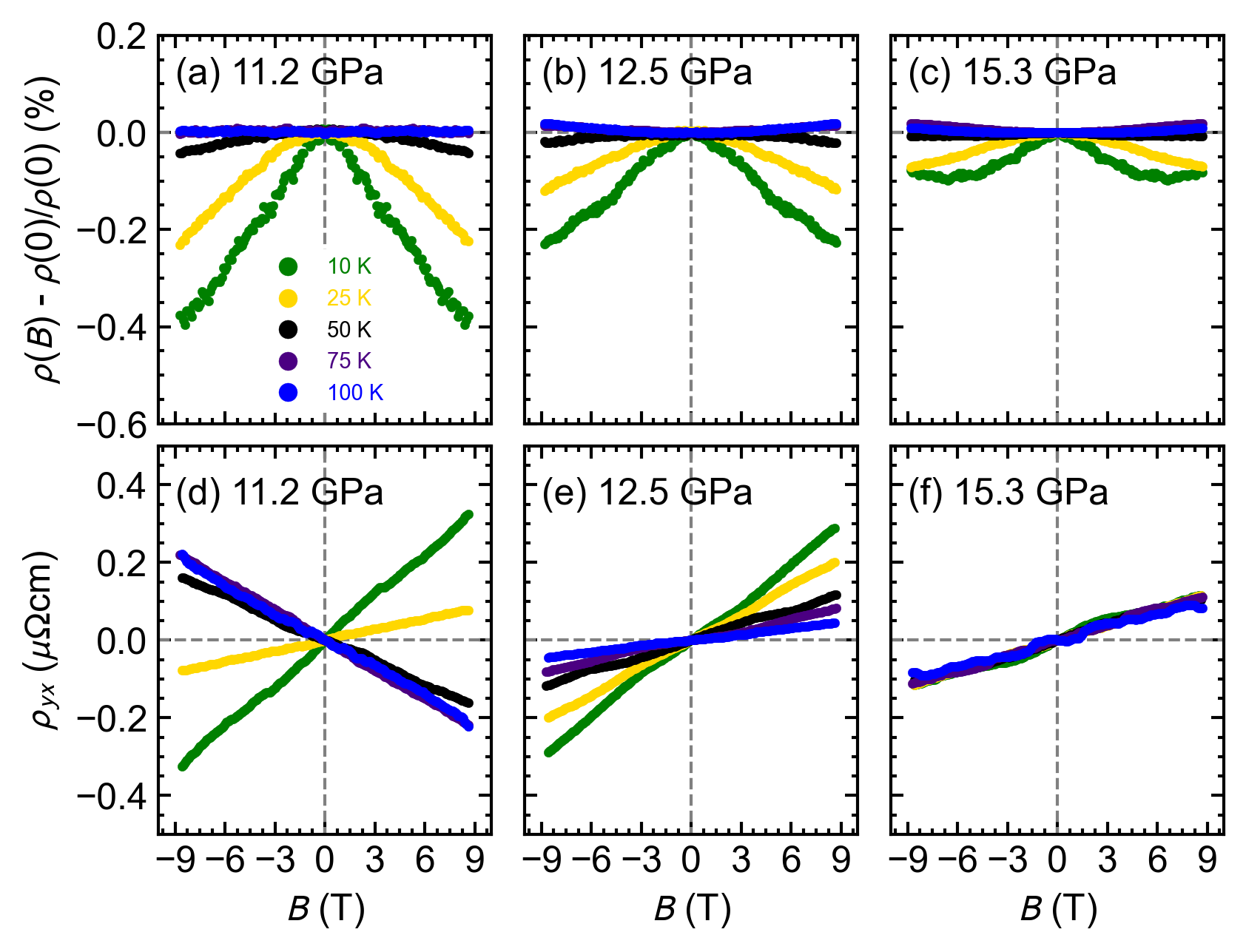}
    \caption{(color online) BaFe$_2$Se$_3$ magnetic field dependencies of (a-c) magnetoresistance and (d-f) Hall resistivity acquired under pressure. (a, d) 11.2 GPa, (b, e) 12.5 GPa, and (c, f) 15.3 GPa.}
    \label{rhoB_BaFe2Se3}
\end{figure}

    Figure \ref{rhoB_BaFe2Se3} shows the magnetoresistance and Hall resistivity of BaFe$_2$Se$_3$ at various pressures. 
    The data were acquired in a transverse magnetic configuration with the current applied in the leg and the magnetic field along the layer. 
    The general features of the magnetoresistance were similar at each pressure; the resistivity decreased with the application of the magnetic field, and its amplitude increased at lower temperatures. 
    This magnetic field dependence was markedly different from the typical features of normal magnetoresistance observed in BaFe$_2$S$_3$. 
    The observed negative magnetoresistance was reminiscent of the magnetoresistance associated with spin fluctuations in (anti-)ferromagnetic metals.
    The negative magnetoresistance was suppressed with increasing pressure, and a positive magnetoresistance was observed near 100 K at pressures above 12.5 GPa.
    This indicated that the antiferromagnetic fluctuations were suppressed toward the $\gamma$ $\rightarrow$ $\delta$ phase transition, suggesting that the normal magnetoresistance was dominant in the $\delta$ phase. 
    
    With regard to the Hall coefficient, because the normal magnetoresistance effect was not observed in BaFe$_2$Se$_3$, we could not perform the analysis based on the two-carrier model.
    Hence, we make a qualitative argument that the sign of the magnetic field dependence of the Hall resistivity was consistent with the sign of the dominant carriers.
    At 11.2 GPa, near the insulator-metal transition, the sign of the Hall coefficient reversed from negative to positive at lower temperatures, indicating that hole carriers were dominant in the normal state near the superconducting phase.
    The temperature dependence of the Hall coefficient diminished with increasing pressure, indicating that the mobility and/or carrier density were no longer temperature dependent. 
    This was consistent with semimetallic electrical resistivity observed on the high-pressure side above the structural phase transition of $\gamma$ $\rightarrow$ $\delta$.

\section{Discussion}
    BaFe$_2$S$_3$ and BaFe$_2$Se$_3$ are quasi-one-dimensional iron-based compounds that exhibit pressure-induced superconductivity and share many features in common.
    The first feature is the pressure at which the insulator-metal transition occurs.
    Despite their 12\% difference in volume at ambient conditions, the critical pressures of the insulator-metal transitions and the superconducting transitions are both near 10 GPa.
    The second feature is that the crystal and magnetic structures of the antiferromagnetic insulating phase adjacent to the superconducting phase are both $\alpha$ and stripe antiferromagnetism.
    The last feature is that, under various pressures, Hall measurements in the normal state near the superconducting phase show that hole carriers were dominant in the metallic phase in both compounds.

    There were also differences in the details between BaFe$_2$S$_3$ and BaFe$_2$Se$_3$.
    The largest difference was the transport properties of the normal state above the superconducting phase.
    BaFe$_2$S$_3$ exhibited a Fermi-liquid-like temperature dependence of electrical resistivity in the normal state and a normal magnetoresistance.
    In contrast, BaFe$_2$Se$_3$ exhibited significant three-dimensional antiferromagnetic fluctuations in the normal phase just above the superconducting phase, as well as a negative magnetoresistance.
    These differences were likely attributed to differences in electronic states. 
    In BaFe$_2$S$_3$, the absence of an antiferromagnetic metallic phase was experimentally confirmed via $\mu$SR and M\"{o}ssbauer measurements~\cite{Zheng2018, Materne2019}.
    In BaFe$_2$Se$_3$, the antiferromagnetic metallic phase remained adjacent to the superconducting phase, as revealed in previous K$\beta$ spectroscopy and neutron powder diffraction experiments~\cite{Ying2017,Zheng2022}.
    The existence of an antiferromagnetic metallic state in BaFe$_2$Se$_3$ was also supported by first-principles calculations~\cite{Zhang2018}.  
    BaFe$_2$Se$_3$ is an important example of coexisting superconductivity and magnetism in orbital-selective Mott insulators.
    The direct observation of the magnetic ordering via $\mu$SR and M\"{o}ssbauer measurements under pressure is also an important issue.

\section{Summary}
    We investigated the electrical resistivity and magnetotransport properties of the iron-based ladder compounds BaFe$_2$S$_3$ and BaFe$_2$Se$_3$ under pressure to reveal the properties of their normal states near the pressure-induced superconducting phase.
    The temperature dependence of the BaFe$_2$S$_3$ electrical resistivity above the insulator-metal transition exhibited a typical Fermi-liquid-like behavior characterized by $\rho$ $\propto$ $T^2$, while that of BaFe$_2$Se$_3$ was characterized by $\rho$ $\propto$ $T^{3/2}$, which is characteristic of three-dimensional spin fluctuations.
    In the magnetotransport measurements, normal magnetoresistance was observed near the superconducting phase in BaFe$_2$S$_3$, while BaFe$_2$Se$_3$ exhibited a negative magnetoresistance that suggested the presence of an antiferromagnetic metallic phase.
    We also investigated the changed BaFe$_2$Se$_3$ electronic structure in the $\gamma \rightarrow \delta$ phase transition via electrical resistivity measurements.
    That transition decreased the electrical resistivity by approximately one order of magnitude, making it almost temperature independent.
    The phase transition thus had a strong impact on electron transport properties despite no symmetry breaking.
    
\begin{acknowledgments}
    We thank Y. Imai, H. Takahashi, and T. Yamauchi for fruitful discussion. 
    This work is financially supported by JSPS KAKENHI Nos.
    JP22H00102,
    JP19H05823, 
    JP19H05822, 
    JP19K21837, 
    JP18H01159, 
    JP16K17732, JP20K14396,
    JP17H05474, JP19H04685
    ;by Murata Science Foundation; and by JST CREST under Grant No. JP19198318.
\end{acknowledgments}

\bibliographystyle{apsrev4-2}
\bibliography{ref}

\end{document}